\documentstyle[times,pramana,epsf,floats]{ias}
\begin{document}
\title{Fluctuation theorems and orbital magnetism in nonequilibrium state}

\author{A.M. Jayannavar and Mamata Sahoo}
\address{Institute of Physics,~~Sachivalaya Marg,~~Bhubaneswar-751005, India}
\abstract{We study  Langevin dynamics of a driven charged particle in the presence as well as in the absence of magnetic field.~~~ We discuss the validity of various work fluctuation theorems using different model potentials and external drives.~~~We also show that one can generate an orbital magnetic moment in a nonequilibrium state which is absent in equilibrium.}

\maketitle
\section{Introduction}
Recent developments in nonequilibrium statistical mechanics has led to the discovery of several rigorous theorems for systems far away from equilibrium[1-10].~~~The fluctuation theorems describe exact relations for properties (symmetries) of distribution functions of various physical quantities such as work,~~heat,~~entropy,~~etc, in the nonequilibrium state.~~~The fluctuation relations are statements about the symmetry of the distributions around zero and not around maximum of physical quantities .~~~They involve negative tails in physical quantities which are usually very rare and are related to transient  second law violating contributions.~~~These theorems are  useful to probe nonequilibrium states in nanophysics and biology.~~~In these systems energies involved are typically small and hence thermal fluctuations play significant role.~~~In fact,~~~ variance in some of the physical quantities dominate the mean value rendering these quantities non-self-averaging.~~~Analyzing the role of these fluctuations may help in understanding and improving the performance characteristics  of engines at nanoscale(e.g,~molecular motors).~~~On the application side,~~~Jarzynski equality[4] has been used to measure equilibrium free energies ($\Delta F$)of the systems from   the statistics of the nonequilibrium work ($W$) performed.
 
There have been an explosion in the number of fluctuation theorems relating various physical quantities in the last few years.~~~Some of these theorems have been verified experimentally on single nanosystems in physical environment  where fluctuations play a dominant role[11,12].~~~~In our present work we study some solvable models[8,9,13] which illustrate the Jarzynski equality and related steady state fluctuation theorems[3,8,9,13,14].~~~We have also studied a driven particle in a nonlinear potential numerically to establish steady state fluctuation theorem for  work .~~~The well- known Bohr-van-Leeuwen theorem states that a classical thermodynamic equilibrium system does not exhibit orbital magnetism[15,16].~~~ However,~~we show that we can obtain the orbital magnetism (paramagnetic/diamagnetic) in  driven nonequilibrium systems. 
\section{The Model}
We consider the dynamics of a charged($e$) Brownian  particle in a two dimensional ($x-y$) plane in the presence of a time dependent potential $U(=U(x,y,t))$.~~~An external magnetic field ($B$) is along $z$ direction.~~~The particle-environment interactions can be treated via Langevin equations[16,17],
\begin{equation}
m\ddot{x} =-\gamma \dot{x}-\frac{\vert{e}\vert}{c}B\dot{y}-\frac{\partial U}{\partial x}+\xi_x(t),
\end{equation}
 \begin{equation}
m \ddot{y}=-\gamma \dot{y}+\frac{\vert{e}\vert}{c}B \dot{x}-\frac{\partial U}{\partial y}+\xi_y(t),
\end{equation}
where the random force field $\xi_{\alpha}(t)$ is a Gaussian white noise with
\begin{equation}
\langle \xi_{\alpha}(t)\xi_{\beta}(t^{'}) \rangle=D\delta_{\alpha \beta} \delta(t-t^{'}) .
\end{equation}
Here $\gamma$ is the friction coefficient and $\alpha,\beta =x,y$.~~~The consistency conditions for the state of equilibrium in the absence of time dependent 
field relates the prefactor $D$ to $\gamma$ as $D=2\gamma k_{B} T$.~~~This problem for time independent potential was considered earlier[15] to elucidate the subtle role played by the boundary conditions in the celebrated theorem of Bohr-van-Leeuwen on the absence of diamagnetism in classical systems[18].~~~This,~~  in turn,~~ implies that free energy of a system is independent of magnetic field.
~~~~Corresponding quantum problem is studied in references [19],~~~with several interesting implications.
Equations (1) and (2) can be written in the over damped regime as 
\begin{equation}
\gamma \dot{x}=-\frac{\vert e \vert B}{c}\dot{y}-\frac{\partial U}{\partial x}+\xi_x(t)
\end{equation}

\begin{equation}
\gamma \dot{y}=\frac{\vert e \vert B}{c}\dot{x}-\frac{\partial U}{\partial y}+\xi_y(t)
\end{equation}
In our following treatment  we consider over damped equations for fluctuation theorems and under damped equations for calculating orbital magnetic moment.

We consider three different protocols for the time dependent potential:~~~ 
(i) Particle in a two dimensional harmonic potential the centre of which is dragged with a uniform velocity in the diagonal direction in x-y plane.~~~For this case $U(x,y,t)=\frac{1}{2}k\vert \vec r-\vec r^{*} \vert^{2}$,~~~~where $\vec r$ is a two dimensional vector ($\vec r=x  \hat i+y \hat j$) and $\vec r^{*}(t)=vt(\hat i+\hat j)$.~~~
(ii) $U(x,y,t)=\frac{1}{2} k (x^{2}+y^{2})-Ax\sin{\omega t}$,~~~~i.e,~~~the particle is subjected to a harmonic ac drive along x direction.~~~For case(iii)
we consider a one-dimensional problem with nonlinear potential 
$U(x)=\frac{1}{4} \alpha x^{4}$ subjected to harmonic drive,~~~i.e,~~~
$U(x,t)=\frac{1}{4}  \alpha x^{4}-Ax\sin{\omega t}$.
\section{Results and Discussions}
Thermodynamic work done on the system for case(i) (or the input energy injected into the system ) for an external agent during a time interval $t$ is given by[8,9,20] 
\begin{equation}
W=-kv\int_0^{t} \{(x(t^{'})-vt^{'})+(y(t^{'})-vt^{'})\} dt^{'} ,
\end{equation}
and for case(ii) and (iii)
\begin{equation}
 W=-A\omega \int_0^t \cos(\omega t^{'})  x(t^{'}) dt^{'} .
\end{equation}
   
It may be emphasized that the thermodynamic work or Jarzynski work is not a mechanical work[21].~~~The thermodynamic work corresponds to the input 
energy pumped into the system by an external time dependent perturbations.~~~~It is clear from the above expressions that we have 
to solve the problem for $x$ formally to obtain work distributions.~~~To solve the problem analytically for the case(i)~~~we define a new variable $z=x+iy$ ($i=\sqrt{-1}$),~~~and with the help of the  over damped equations (4) and (5),~~~we get

\begin{equation}
\dot{z}=\frac{-kpz}{\gamma}+\frac{kpg^{*}(t)}{\gamma}+\frac{p\xi(t)}{\gamma} ,
\end{equation}
where $p=\frac{1+iC}{1+C^{2}}$, $\xi(t)=\xi_{x}(t)+i\xi_{y}(t)$,~   
$g^{*}(t)=vt(1+i)$ and $C=\frac{e \vert B \vert}{\gamma c}$.\\

The formal solution for equation (8) is given by
\begin{eqnarray}
z(t)&=&z_{0} \exp(-\frac{k}{\gamma}pt)+\frac{p}{\gamma} \int_0^t dt^{'} \exp(-\frac{k}{\gamma}p(t-t^{'}))\{kg^{*}(t^{'})+\xi(t^{'})\},
\end{eqnarray}
where $z_{0}=x_{0}+iy_{0}$,~~~~and $x_{0}$ and $y_{0}$ are initial co-ordinates  of the particle at time ~~$t=0$.~~~~It may be readily noticed from equation (6) that particle co-ordinates at time $t$  and consequently  work done are linear functionals of Gaussian variables.~~~Hence it follows that work distribution[8,9,13] is a Gaussian ,~~~which can be completely specified by mean$\langle W \rangle$ and the variance $\sigma^{2}=\langle W^{2} \rangle -\langle W \rangle^{2}$.~~~~It is straight forward to calculate these quantities.~~For this we refer to [13].~~~~Final result for $\langle W \rangle $ is given by 

\begin{eqnarray}
\langle W \rangle &=& 2 \gamma v^2 \{t-\frac{\gamma}{k}(1-\exp(-k^* t)
\cos(\Omega t))-\frac{C \gamma}{k}\sin(\Omega t)\nonumber\\&&\exp (-k^* t)\}-\gamma v^2 2 C\{\frac{\gamma}{k}\sin(\Omega t)\exp(-k^*t)-\frac{C \gamma}{k}\nonumber\\&&(1-\exp(-k^*t)\cos(\Omega t))\} ,
\end{eqnarray}
where
$\Omega = \frac {k C}{\gamma(1+C^2)}$ and 
$k^{*}=\frac{k}{\gamma(1+C^{2})}$.\\

The variance of the work 
\begin{equation}
\sigma^{2}=\langle W^2 \rangle -\langle W \rangle^2 = \frac{2 \langle W \rangle }{\beta} ,
\end{equation}
Here $\beta=\frac{1}{k_{B}T}$.~~~~
To obtain the above  results (eqns (10) and (11)) we have assumed initial distribution for the  co-ordinates $x_{0}$ and $y_{0}$ to be equilibrium distribution,~~
$P_{e}(x_{0},y_{0},t)=\frac{\beta k}{2\pi} \exp[\frac{-\beta k (x_{0}^{2}+y_{0}^{2})}{2}]$.~~~~The full probability distribution $P(W)$ is 

\begin{eqnarray}
P(W)&=&\frac{1}{\sqrt{4\pi\langle W \rangle/\beta}} 
e^{-(W-\langle W \rangle)^{2}/(4\langle W \rangle/\beta)} .
\end{eqnarray}
 
The Jarzynski equality follows immediately namely

\begin{equation}
\langle e^{-\beta W} \rangle= e^{-\beta \Delta F}=1 .
\end{equation}

The above equation implies $\Delta F=0$,~~~indicating that the equilibrium free energy difference ($\Delta F$) is independent of magnetic field consistent with Bohr-van-Leeuwen theorem.~~~Jarzynski equality relates nonequilibrium quantities with equilibrium free energies.~~~Initially the system is assumed to be in equilibrium defined by a thermodynamic parameter $A$ (in our present case centre of the harmonic potential ).~~~The nonequilibrium process is obtained by changing the thermodynamic control parameter with a prescribed protocol up to time $\tau$,~~~where the thermodynamic parameter has value $B$.~~~The state of the system at the end of the protocol is not in equilibrium.~~~This protocol is repeated for large number of times.~~~For each realization we get a different quantity $W$.~~~ Using eqn.~(1),~~~one obtains free energy difference $\Delta F=F_{B}-F_{A}$ after evaluating the average $\langle ... \rangle$ over all possible realizations.~~~~In our present case the free energy is independent of the centre of the harmonic oscillator and the applied  magnetic field and hence $\Delta F=0$.~~~~However,~~~it may be noticed that the thermodynamic work (eqn.~(10)) depends on the magnetic field and there is a finite probability of $W$ being negative.~~~~The relaxation rate $\tau_{r}(=\frac{\gamma(1+C^{2})}{k})$  also depends on the magnetic field.~~~In the absence of magnetic field we reproduce the results obtained in the references[8,9].~~~~Discussion of the above distribution for $P(W)$ in the asymptotic time limit $t \rightarrow \infty $ ($t \gg \tau_{r}$) in connection with steady state fluctuation theorem and Hatano-Sasa identity is discussed in ref[13].

We now turn to case(ii)  to examine the steady state fluctuation theorem .~~~~In the large time regime probability distributions are time periodic with a period ($\frac{2\pi}{\omega}$).~~~~The problem being linear we can calculate average work done $\langle W \rangle$ and variance over a single  period ( $\frac{2\pi}{\omega}$)  analytically as given by 
\begin{equation}
\langle W_{s} \rangle=\lim_{t \rightarrow \infty}[\langle W(t+\frac{2\pi}{\omega})\rangle-\langle W(t)\rangle]
\end{equation}

\begin{equation}
   =\frac{\pi A^2  \omega \gamma(k^2+\omega^2\gamma^{2}(1+C^2))}{(k^2+(1+C^2)\gamma^{2}\omega^2)^2-4k^2C^2 \gamma^{2}   \omega^{2}}
\end{equation}

\begin{equation}
\langle V_{s} \rangle = \langle W_{s}^{2} \rangle-\langle W_{s} \rangle^{2}
\end{equation}
\begin{equation}
\langle V_{s} \rangle =\frac{2}{\beta}\langle W_{s} \rangle
\end{equation}

The probability distribution of $W_{s}$ is again Gaussian and satisfies the relation

\begin{equation}
\frac{P(W_{s})}{P(-W_{s})}=e^{\beta W_{s}}
\end{equation}

The above equation is a statement[8,9,13,14] of steady state fluctuation theorem[SSFT].~~~~Thus we have shown that work done over a single period in the time asymptotic periodic regime satisfies SSFT.~~~~We would like to emphasize that the validity of SSFT over a single period is restricted only to over damped linear models.~~~~In general this will not hold true in nonlinear situations.~~~~However,~~we will show later that SSFT holds even for nonlinear models  if one considers the work done over a large number of periods or over a single period ,~~~however in the large noise limit.~~~The convergence of SSFT on accessible time scales has been discussed in the previous literature [14].~~~It may also be noted that the average work done over a period (en 15) depends on magnetic field.~~~~However,~~~it is independent of temperature which is again valid for a considered over damped linear model only.

To discuss the validity of SSFT in over damped nonlinear systems we turn to case(iii) where the particle in quartic potential is subjected to an ac force .~~~~In the absence of magnetic field it reduces to a one dimensional problem.~~~~Numerical simulations of this model was carried out by using Heun's method[22].~~~~To calculate the  work done over a period (eqn. 7) in the time asymptotic regime we neglect initial transients and work done over  a period is calculated.~~~~To get better statistics we have calculated $W_{s}$ over more than 1000000 realizations.  

\begin{figure}[htbp]
\epsfxsize=11cm
\centerline{\epsfbox{fig1a.eps}}
\label{fig:food}
\vspace{1.4cm}
\epsfxsize=11cm
\centerline{\epsfbox{fig1b.eps}}
\label{fig:food}
\end{figure}
\vspace{0.1cm}
\begin{figure}[htbp]
\epsfxsize=11cm
\centerline{\epsfbox{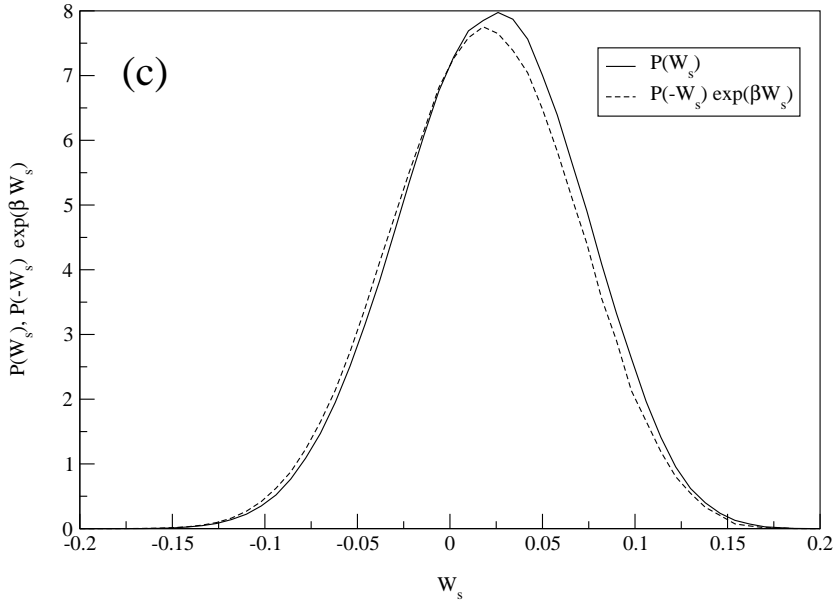}}
\label{fig:food}
\vspace{1.4cm}
\epsfxsize=11cm
\centerline{\epsfbox{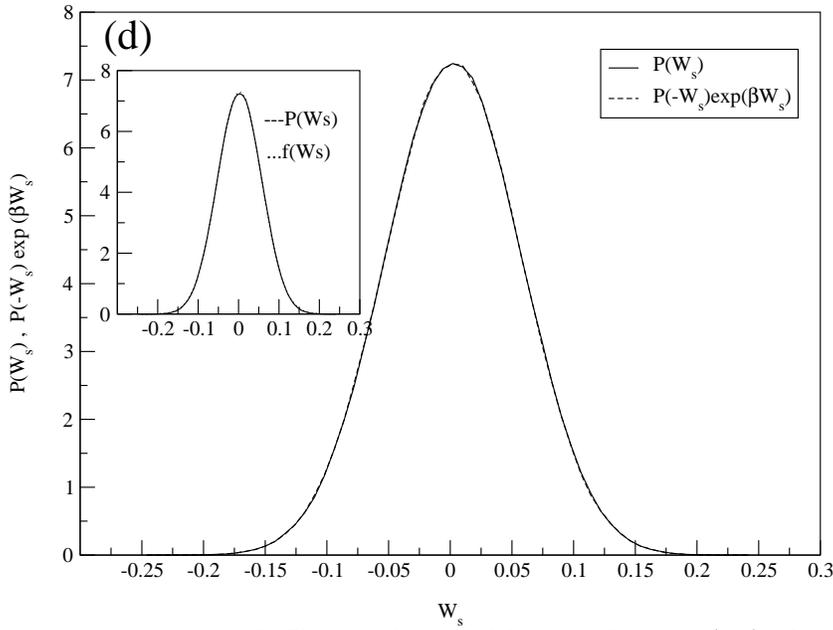}}
\label{fig:food}
\caption{(a-d) Plot of the probability distribution $P(W_{s})$ along with $P(-W_{s}) \exp(\beta W_{s})$ for different noise strengths $D=0.02,0.04,0.06 ,0.6 $ respectively in figs (a),(b),(c) and (d) .~~~Other parameters are $A=0.1$,~~ $\omega=0.1$ and $\alpha=1$.~~In the inset of (d), $f(W_{s})$ is a Gaussian function.  }
\end{figure}

In figs (1-a)-(1-d) we have plotted $P(W_{s})$ and $P(-W_{s}) \exp(\beta W_{s})$ as a function of $W_{s}$ over a single period of the ac force for different values of temperature or the noise strength ($D=0.02,~~0.04,~~0.06,~~0.6$).~~~~All the physical parameters are in dimensionless units and their values are mentioned in the figure captions.~~~~We observe that for small values of noise strength SSFT does not hold,~ i.e.,~~$P(W_{s}) \neq P(-W_{s}) exp(\beta W_{s})$.~~~~However,~~only in the large noise limit (see fig 1-d) SSFT is indeed satisfied within our numerical accuracy.~~~The large noise (or temperature) limit corresponds to a case where the  relaxation time($\sim \sqrt(\frac{\gamma^{2}}{\alpha k_{B}T}$)) in the system becomes much less than a given period of ac force.~~~~The same conclusions can be drawn if one starts in a low noise regime.~~~~However,~~~for this case one has to evaluate the work done over a large number of cycles such that the relevant relaxation time becomes much less than the total time over which the work distribution is evaluated.~~~~~In both the above cases total work can be treated as an addition of independent increments (each increment corresponds to work done over a relaxation time).~~~~Then the central limit theorem leads us to expect that the distribution of work will be Gaussian.~~~~And this is,~~indeed,~~the case.~~~We observe that for $D=0.6$,~~  $P(W)$ approaches a Gaussian  
distribution (In the inset of fig (1-d)  we have shown a Gaussian fit where the mean and variance are calculated from the numerical data).~~~~For this distribution the variance $V$ and the mean $\langle W \rangle$ are related by the fluctuation dissipation ratio $\frac{2 W}{V \beta}=1$,~~so as to satisfy SSFT.~~~~In our case,~~~for $D=0.6$ this ratio is $\approx$ 0.98.~~~~~A similar conclusion was arrived at recently on work fluctuations in systems exhibiting stochastic resonance[23].~~~~Here it is shown that work distribution satisfies SSFT provided one considers  work done over a large number of periods (low temperature regime mentioned earlier) and in this case distribution approaches Gaussian. 

In all the cases studied above we observe that there is always a weight towards negative values of $W_{s}$.~~~~The negative values of $W_{s}$ 
corresponds to the transient second law violating hysteresis loops.~~~~In the time periodic asymptotic state work done over the periods is dissipated into the system as heat[8].~~~~Thus one can identify $\langle W_{s} \rangle $ as hysteresis loss (heat) in the medium.~~~~However,~~it may be noted that fluctuations of the work can not be identified with heat fluctuations[8].~~~~In fig(2) we have plotted the probability distributions of work $W$,~~~~$P(W)$,~~~ that of change in internal energy $\Delta U$,~~ $P(\Delta U)$ and that of  heat $Q$, $P(Q)$.~~~~All these physical quantities  are averaged over a time interval of a single period for $D=0.6$.~~~~From the first law of thermodynamics it follows that $W=Q+\Delta U$.~~~$P(\Delta U)$ is symmetric ($\langle \Delta U \rangle =0$) and the distribution is exponential,~~ i.e.,~~~$P(\Delta U) \sim e^{-\beta \vert \Delta U \vert}$.~~~~As mentioned earlier the distribution $P(W)$ is Gaussian.~~~~At large $Q \gg \langle Q \rangle $ the distribution $P(\Delta U)$ dominates over a Gaussian distribution of $P(W)$ and hence it follows that $P(Q) \sim e^{-\beta \vert Q-\langle Q \rangle \vert}$.~~~~The presence of this exponential tail in $P(Q)$ ,~~~~a new extended fluctuation theorem for heat is obtained[8].~~~~The consequence of this new theorem for heat fluctuations  is that the ratio of the probability for  the Browinian particle to absorb  and to supply heat to the environment is much larger than the one corresponding to the conventional SSFT for the work.~~~Details of the fluctuation theorem for the heat in our nonlinear models will be published elsewhere.~~~~We have also observed that the average hysteresis loss over a period monotonically decreases with temperature as opposed to the case of linear model where it is independent of temperature.
\vspace{1cm}
\begin{figure}[htbp]
\epsfxsize=11cm
\centerline{\epsfbox{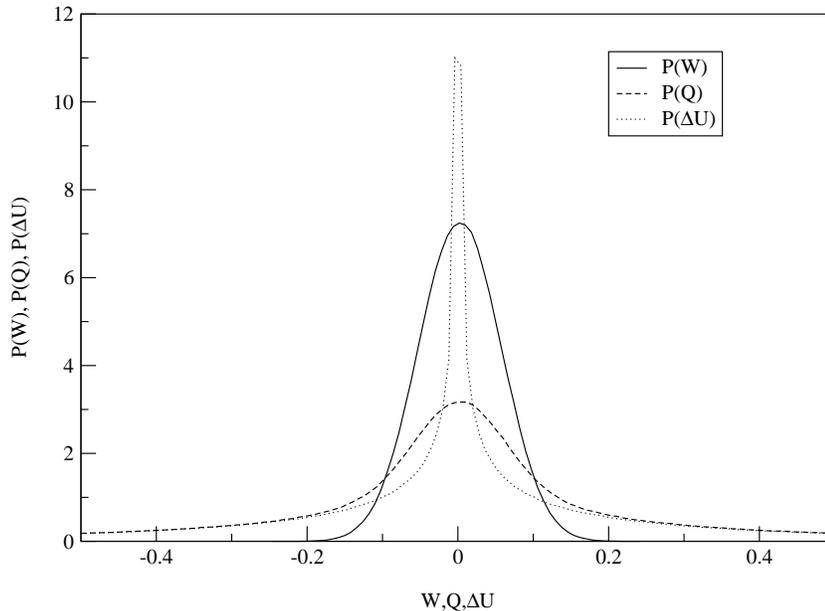}}
\label{fig:food}
\caption{Probability distributions for $W$, $Q$ and $\Delta U$ for temperature $D=0.6$.~~~The other parameters being the same as in fig(1).}
\end{figure}

We finally discuss the problem of orbital magnetism and hysteresis loss in the inertial regime.~~~~For this we use Langevin equations with inertia (eqns (1) and (2)).~~~~The potential considered here is $U(x,y,t)=\frac{1}{2} k (x^{2}+y^{2})-Ax\sin{\omega t}$ (case-ii).~~~~In the time asymptotic regime one can readily obtain expressions for the averaged work done over a single period $\langle W_{s} \rangle $ as well as averaged magnetic moment of the system over a period,~~ namely[15],~~ 
\begin{equation}
\langle M \rangle=\lim_{t \rightarrow \infty}[-\frac{\vert e \vert}{2mc} \frac{\omega}{2\pi}\int_t^{t+\frac{2\pi}{\omega}} \langle \vec{r} \times \vec{v} \rangle dt^{'}],
\end{equation}
where $\vec r$ and $ \vec v$ are the two dimensional position and velocity  respectively.~~~~We obtain, 
\begin{eqnarray}
\langle M \rangle=\frac{-\frac{\vert e \vert}{2mc} (\frac{A}{m})^{2} \omega_{c} \omega^{2}(\omega^2-\Omega^2)}{D}
\end{eqnarray}
where $\omega_{c}=\frac{\vert e \vert B }{mc}$  the cyclotron frequency,~~~$\Omega=\sqrt{\frac{k}{m}}$ is the natural frequency of the harmonic oscillator and  $\Gamma=\frac{\gamma}{m}$.~~~~The hysteresis loss per period $\langle W_{s} \rangle$ is given by 
\begin{eqnarray}
\langle W_{s} \rangle&=&\frac{\pi \frac{A^{2}}{m} \Gamma \omega \big[(\Omega^{2}-\omega^{2})^{2}+(\omega_{c}^{2}+\Gamma^{2})\omega^{2}\big]}{D}
\end{eqnarray}
where 
\begin{eqnarray}
D&=&\big[(\omega_{c}^{2}+\Gamma^{2})^{2}\omega^{4}+(\Omega^{2}-\omega^{2})^{2}\{2\omega^{2} (\Gamma^{2}-\omega_{c}^{2})+(\Omega^{2}-\omega^{2})^{2}\}\big]
\end{eqnarray}
\vspace{0.3cm}

\begin{figure}[htbp]
\epsfxsize=11cm
\centerline{\epsfbox{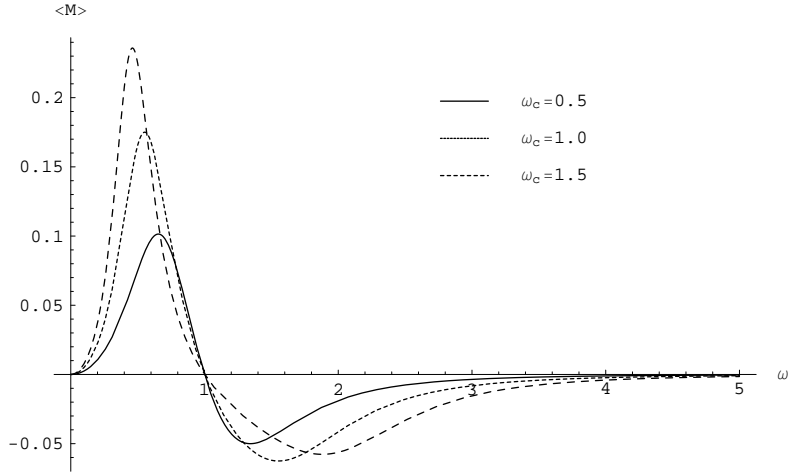}}
\label{fig:food}
\caption{The averaged magnetic moment $\langle M \rangle $ with frequency $\omega$ for different values of $\omega_{c}$,~~for  $k=1$,~$m=1$,~$A=1$ and $\gamma=1$.}
\end{figure}
\vspace{0.3cm}

\begin{figure}[htbp]
\epsfxsize=11cm
\centerline{\epsfbox{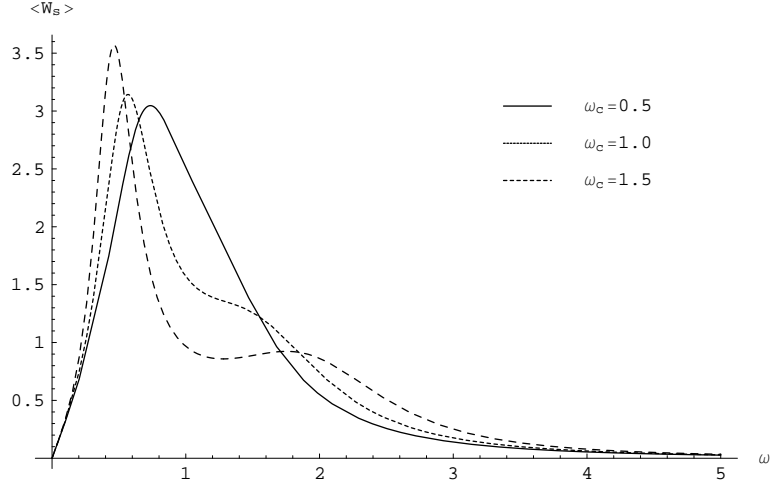}}
\label{fig:food}
\caption{Figure shows the averaged hysteresis loss $\langle W_{s} \rangle$ over one period with frequency $\omega$ for different values of $\omega_{c}$.~~Other parameters are same as in fig(3).}
\end{figure}

In fig(3) we have plotted dimensionless magnetic moment ($\langle M \rangle \equiv (\frac{\langle M  \rangle c \gamma^{3}}{e A^{2}})$) as a function 
of 
dimensionless frequency $\omega$ for three different values of cyclotron frequencies.~~~~The frequencies are scaled with respect to $\Gamma$.~~~In fig(4) we have plotted hysteresis loss in dimensionless form ($\langle W_{s} \rangle \equiv (\frac{\langle W_{s} \rangle m \Gamma^{2}}{A^{2}})$)  as a function of frequency $\omega$ for different values of $\omega_{c}$.~~~~We have set $A$,~~$\gamma$  and $m$ to unity.
 
From the calculation of  $\langle M \rangle$ we state some  noteworthy 
observations.~~~(i)~~~In a nonequilibrium state we obtain magnetic moment,~~~this does not violate Bohr-van Leeuwen theorem as it is valid only  for systems in equilibrium.~~~~(ii)~$\langle M \rangle$ goes to zero as $\omega \rightarrow  0$,~~~ this limit corresponds to equilibrium state.~~~(iii)~~~For small $\omega$,~~~ $\langle M \rangle$ is paramagnetic  and crosses over to a  diamagnetic regime at resonance frequency $\omega=\sqrt{\frac{k}{m}}$  and $ \langle M \rangle \rightarrow 0$ as $\omega \rightarrow \infty$.~~~(iv)~ In both paramagnetic and diamagnetic regime $\langle M\rangle$ exhibits a peak  as a function of $\omega$.~~~~(v)~~~As $B \rightarrow 0$, ~~~$\langle M \rangle \rightarrow 0$ and again $\langle M \rangle$ goes to zero in the large field limit exhibiting a peak as a function of magnetic field $B$  for a nonzero fixed $\omega$.~~~~(vi) ~~~As a function of friction $\Gamma$,~~~~
$\langle M \rangle $ decreases monotonically and goes to zero  as $\Gamma \rightarrow \infty $.~~~~These results are expected on physical grounds.
~~~Hysteresis area exhibits a double peak behavior as a function of the frequency $\omega$ in the inertial regime.~~~~This is not the case for the over damped motion (see eqn.(15)), in the absence of magnetic field ($C=0$).~~~These complex structures are attributed to interplay in dynamics with different frequencies,~~~namely,~~~ cyclotron 
frequency $\omega_{c}$,~~~natural frequency $\Omega$ along with frequency $\omega$ of a forcing.

It may  also be noted that in the absence of magnetic field systems do not possess  angular momentum.~~~However,~~~~ if we apply two oscillating ac fields along $x$ and $y$ direction respectively(in absence of magnetic field) the system can acquire  angular momentum  as the resultant field drags the particle in circular or elliptical orbit depending on the relative strength and phase difference between the two perpendicular ac fields.~~~~We have also solved this problem analytically for $\langle M \rangle $ as well as the work distributions.~~~~ Our results for work fluctuation  in the inertial regime indicate that $P(W)$ is Gaussian and satisfies SSFT.
\section{Conclusion}
By studying the dynamics of a charged particle in the presence of magnetic field in nonequilibrium state we have verified Jarzynski equality.~~~In particular we have shown that nonequilibrium state supports orbital magnetism without violating Bohr-van-Leeuwen theorem.~~~Steady state fluctuation theorem for work is discussed for two different model systems.~~~In the regime of validity of SSFT we have observed that distribution for work approaches a Gaussian distribution.
\section{Acknowledgements}
One of us (AMJ) thanks Abhishek Dhar  and N.Kumar for helpful discussions.

\end{document}